\documentclass[12pt,english]{article}

\usepackage{amsmath,amssymb }
\usepackage{fancybox}
\usepackage{graphicx,psfrag,epsf}
\usepackage{enumerate}
\usepackage{graphicx,psfrag}
\usepackage{multirow}
\usepackage{epsfig}
\usepackage{subfigure}
\usepackage{theorem}
\usepackage{natbib,psfrag}
\usepackage[usenames,dvipsnames]{color}
\usepackage{todonotes}
\usepackage{xcolor}
\usepackage{algorithm2e}
\usepackage{amsfonts}
\usepackage{booktabs}
\usepackage{siunitx}

\newcommand{\pdf}{0}
\if1\pdf
\usepackage[pdftex]{graphicx}
\else
\usepackage{graphicx}
\fi

\newcommand{\blind}{1}

\def\mb#1{\setbox0=\hbox{$#1$}
  \kern-.025em\copy0\kern-\wd0
  \kern.05em\copy0\kern-\wd0
  \kern-.025em\raise.0em\box0}

\makeatletter
\newcommand*\dashline{\rotatebox[origin=c]{90}{$\dabar@\dabar@\dabar@$}}
\makeatother

\newcommand{\pr}{\text{Pr}}
\newcommand{\cov}{\text{Cov}}

\newcommand{\diag}{\text{diag}}

\begin{document}

\baselineskip 1.8em
\parskip 1em

%
%

\def\spacingset#1{\renewcommand{\baselinestretch}%
	{#1}\small\normalsize} \spacingset{1}

\if1\blind
{
	\title{\bf Adaptive Bayesian Power Spectrum Analysis of Multivariate Nonstationary Time Series}
	\author{Zeda Li and Robert T. Krafty
		\footnote{Z. Li is PhD student, Department of Statistical Science, Temple University (zeda.li@temple.edu) and R. T. Krafty is Associate Professor, Department of Biostatistics, University of Pittsburgh (rkrafty@pitt.edu).   This work was supported by NIH grants R01GM113243 and R01HL1104607.  The authors thank two Referees, an Associate Editor and Editor for providing insightful comments that greatly improved the article.}}
	\maketitle
} \fi

\if0\blind
{
	\bigskip
	\bigskip
	\bigskip
	\begin{center}
		{\LARGE\bf Title}
	\end{center}
	\medskip
} \fi

\newpage

\setlength{\baselineskip}{22pt}  

\begin{center}
\section*{Abstract}
\end{center}
This article introduces a nonparametric approach to multivariate time-varying power spectrum analysis. The procedure adaptively partitions a time series into an unknown number of approximately stationary segments, where some spectral components may remain unchanged across segments, allowing components to evolve differently over time.
Local spectra within segments are fit through Whittle likelihood based penalized spline models of modified Cholesky components, which provide flexible nonparametric estimates that preserve positive definite structures of spectral matrices.
The approach is formulated in a Bayesian framework, in which the number and location of partitions are random, and relies on reversible jump Markov chain and Hamiltonian Monte Carlo methods that can adapt to the unknown number of segments and parameters. By averaging over the distribution of partitions, the approach can approximate both abrupt and slow-varying changes in spectral matrices.   Empirical performance is evaluated in simulation studies and illustrated through analyses of electroencephalography during sleep and of the El Ni\~{n}o-Southern Oscillation.

\noindent%
KEY WORDS: Locally Stationary Process; Modified Cholesky Decomposition; Nonstationary Multivariate Time Series; Reversible Jump Markov Chain Monte Carlo; Penalized Splines; Spectral Analysis.

\bigskip

\section{Introduction}
\begin{figure}[t]
	\centering
	\includegraphics[height=2.1in]{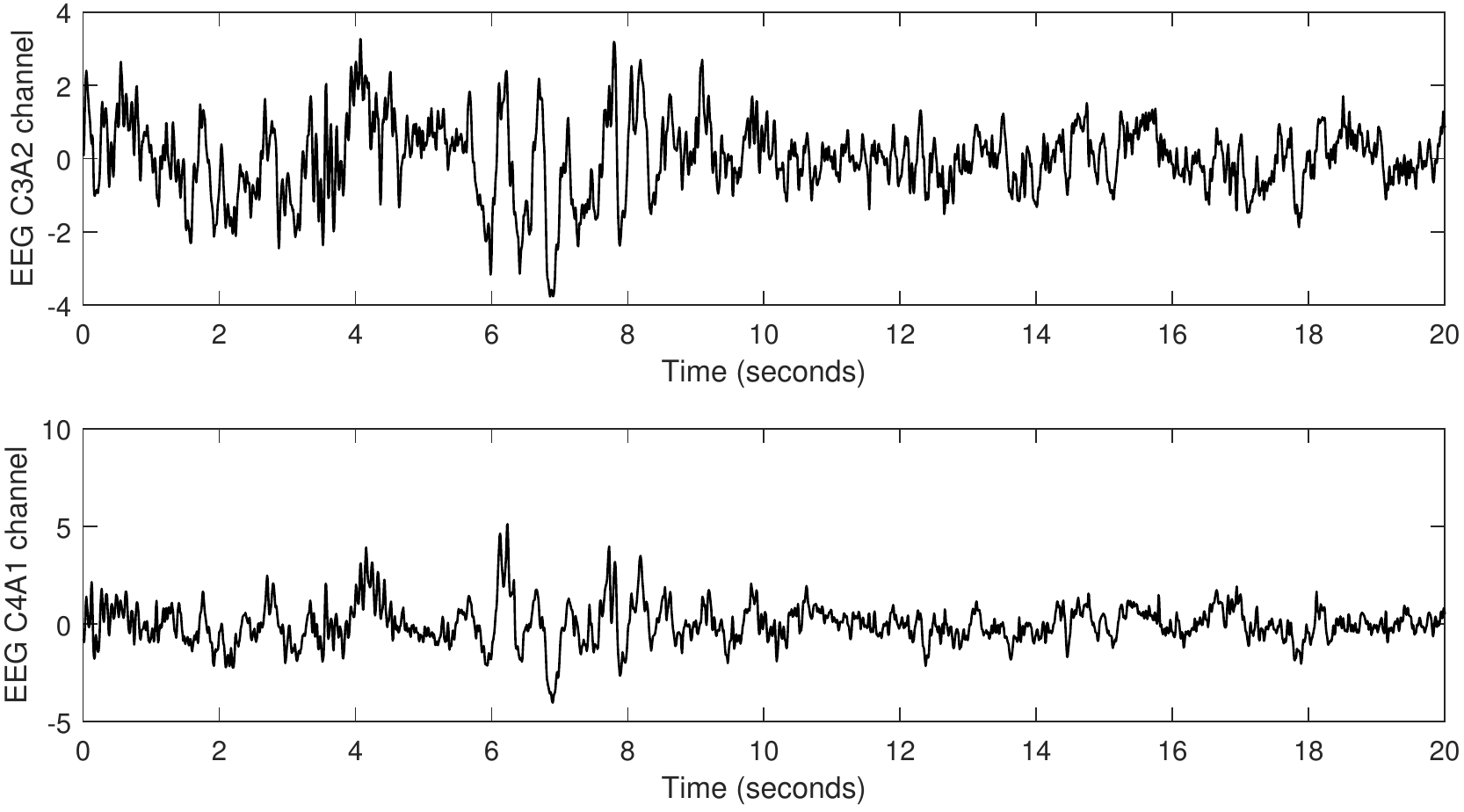}
	\caption{Two-channel EEG from a patient during sleep. }
	\label{tseeg}
\end{figure}

Understanding second-order frequency domain properties of multivariate time series is essential to addressing scientific questions in a variety of fields.
The time series of interest are often nonstationary and the manner in which frequency domain properties evolve over time can provide important scientific information.  An example of such data comes from sleep medicine, where physicians commonly record patients' brain activity during sleep via multi-channel electroencephalography (EEG) to inform the treatment of sleep disorders.  Brain activity is dynamic, so that the data are inherently nonstationary, and the frequency domain properties of EEG provide interpretable biological information.  An example of an epoch of two-channel EEG is displayed in Figure \ref{tseeg}.
%

Methods for estimating the time-varying spectrum of a multivariate time series
can be roughly grouped into three categories.  The first category consists of estimators in which second-order frequency domain structures evolve continuously over time.  This includes the slowly-evolving multivariate locally stationary process,  which can be analyzed parametrically by fitting time-series models with time-varying parameters \citep{dahlhaus2000}, and nonparametrically via the bivariate smoothing of spectral components as functions of frequency and time \citep{guo2006}.     Additional nonparametric methods based on locally stationary wavelet models have been  developed that, although also assume that second-order structures evolve continuously, are better localized in time and frequency to capture more rapid but continuous changes \citep{sanderson2010, park2014}. The second category of methods consist of estimators that are piecewise stationary.  Approaches within this second category typically divide a time series into approximately stationary segments, then obtain estimates of local spectra within segments.  These methods include both  parametric approaches, such as fitting piecewise vector autoregressive  models
\citep{davis2006}, and nonparametric approaches, such as using the multivariate smooth localized complex exponential (SLEX) library
\citep{ombao2005}.   The final category are methods that can automatically approximate both abrupt and slowly-varying changes by averaging over piecewise stationary models, and include the stochastic approximation Monte Carlo (SSAMC) based method of \cite{zhang2016}.

The property that spectral matrices must be non-negative definite makes the time-varying spectral analysis of a multivariate time series a challenging problem.  In particular, to assure that estimates are positive definite, most existing approaches require a common temporal structure across spectral components.  For piecewise stationary based estimates, this equates to all spectral components having the same partitions, and inhibits the accurate modeling of processes where one component, say the cross-spectrum between two series, changes while other components, such as the individual spectra of the two series, remain stationary.  An exception to this limitation is the method of \cite{guo2006}, which models Cholesky components of spectral matrices to assure estimates are positive definite.  However, their approach is highly dependent on slowly-varying temporal smoothness and cannot capture rapid or abrupt changes.


The goal of this article is to expand the scope of processes that can be accurately analyzed by introducing a novel multivariate time-varying spectral analysis procedure that can capture both abrupt- and smoothly-changing spectral dynamics while flexibly allowing components to evolve differently in both time and frequency.   The proposed procedure adaptively approximates a multivariate locally stationary time series through piecewise stationary processes with a random number of segments and partitions.  Using reversible jump Markov chain Monte Carlo (RJMCMC) techniques with Hamiltonian Monte Carlo (HMC) updates conditional on the number of partitions, the approach estimates the unknown number and location of partitions and fits penalized splines to the modified Cholesky components of local spectral matrices.  By modeling modified Cholesky components, the method is able to allow some components to remain stationary across a partition point while producing positive definite estimates.  The sample generated from the sampling algorithm can be used to investigate the number and location of change points and, by averaging estimates over the distribution of partitions, not only produces estimates that can capture abrupt changes, but also effectively approximates slowly-varying dynamics.
To the best of our knowledge, this is the first methodology for multivariate spectral analysis that flexibly allows spectral components to evolve differently over time while capturing both abrupt and slowly-varying dynamics.
In addition to allowing components to evolve differently over time, empirical examples suggest that the proposed approach is computationally faster and can produce more accurate estimates compared to
other approaches that can approximate both abrupt and slowly-varying dynamics \citep{zhang2016}.


The rest of this article is as follows.  A model for time-varying multivariate power spectra is introduced in Section \ref{sec:model}.  In Section \ref{sec:prior}, prior distributions for model parameters are formulated  and a sampling scheme is outlined to provide a fully automated analysis.  Simulation studies are presented in Section \ref{sec:example} to illustrate the proposed procedure's ability to capture both slow-varying and abrupt changes, to estimate the number and location of partitions,  and to compare its performance to that of other estimators.
The use of the procedure in real-world applications is illustrated in Section \ref{sec:app}, where it is used to analyze EEG from a patient during sleep and to analyze the El Ni\~{n}o-Southern Oscillation. Concluding remarks are given in Section \ref{remarks}.  A detailed description of the sampling scheme is given in the supplementary materials.

\section{The Model}\label{sec:model}

\subsection{Time-Varying Spectrum}\label{subsec:locally}
This article considers the time-varying spectral analysis of a locally stationary $N$--dimensional time series defined through a Cram\'{e}r representation with time-varying transfer function.  The use of time-varying transfer functions to generalize power spectra for  stationary time series to the nonstationary setting was introduced by \citet{priestley1965} for univariate time series, and extended to multivariate time series by \cite{dahlhaus2000} and \citet{guo2006}.   A time-varying transfer function $A(u, \omega)$ is a function of scaled time $u \in [0,1]$ and frequency $\omega \in \mathbb{R}$ such that $A(u, \omega)$ is a nonsingular $N \times N$ complex valued matrix that is periodic and Hermitian as a function of frequency, or $A(u, \omega) = A(u, \omega + 2 \pi)$ and $A(u, \omega) = A(u, -\omega)^*$, where $A^*$ is the complex conjugate of $A$.  Formally, we consider $\mathbb{R}^N$--valued time series of length $T$, $\left\{ \mb X_t : t =1, \dots, T \right\}$, of the form
\begin{equation*}\label{eq:X}
\mb X_t = \int_{-1/2}^{1/2} A(t/T, \omega) \exp{(2 \pi i \omega t)}  d \mb Z(\omega),
\end{equation*}
for a time-varying transfer function $A(u, \omega)$ and an $N$--dimensional mean-zero orthogonal process $\mb Z(\omega)$  that is Hermitian and where $E\left\{  d \mb Z(\omega) d \mb Z^*(\zeta) \right \}$ is the identity matrix if $\omega = \zeta$ and zero otherwise.

The focus of this article is on estimating the time-varying spectrum
\begin{equation*}
f(u, \omega) = A(u, \omega)A(u, \omega)^*, \quad u \in [0,1], \omega \in \mathbb{R}.
\end{equation*}
The time-varying spectrum $f(u, \omega)$ is a  positive definite Hermitian $N \times N$ matrix. We assume it satisfies some regularity conditions both as a function of $\omega$ and as a function of $u$.  For every $u$, we assume that each component of $f(u, \cdot)$  possesses a square-integrable first derivative as a function of frequency.  For every $\omega$, we assume that each component of $f(\cdot, \omega)$ is continuous as a function of scaled time at all but a possible finite number of points.

We refer to this class of time series as locally stationary and note that it differs somewhat from the locally stationary time series models considered by \cite{dahlhaus2000} and \citet{guo2006}.  First, \cite{dahlhaus2000}  assumed a series of transfer functions $A^0_{t,T}(\omega)$ that converge to a large sample transfer function $A(u, \omega)$ in order to allow for the fitting of parametric models.  Since we are considering nonparametric estimation, in a manner similar to \citet{guo2006}, we define our model directly using $A(u, \omega)$.  Second, the models of \cite{dahlhaus2000} and \citet{guo2006} require the time-varying spectrum to be continuous in both time and frequency.  The proposed model is more flexible and allows for components of spectral matrices to evolve not only continuously, but also abruptly in time.

\subsection{Piecewise Stationary Approximation}
Our analysis of a locally stationary time series $\left\{ \mb X_t : t =1, \dots, T \right\}$ begins by noting that it can be well approximated by a piecewise stationary process \citep{adak1998, guo2006}.  Consider a partition of the time series into $m$ segments defined by partition points $\mb \delta = \left(\delta_{0}, \dots, \delta_{m}\right)'$  with $\delta_{0} = 0$ and $\delta_{m} = T$ such that $\mb X_t$ is approximately stationary within the segments $\left\{t : \, \delta_{q-1} < t \le \delta_{q}\right\}$ for $q=1, \dots, m$.  Then
\begin{equation*}
\mb X_t \approx \sum_{q=1}^{m} \int_{-1/2}^{1/2} A_q(\omega) \exp{(2 \pi i \omega t)} d \mb Z(\omega),
\end{equation*}
where $A_q(\omega)=A(u_q, \omega) I(\delta_{q-1} < t \le \delta_q)$, $I(\cdot)$ is the indicator function, and $u_q = \left( \delta_{q} + \delta_{q-1}\right)/2T$ is the scaled midpoint of the $q$th segment.
Within the $q$th segment, the time series is approximately second-order stationary with local power spectrum $f(u_q, \omega)= A_q(\omega)A_q(\omega)^*$.

Conditional on an approximately stationary partition $\mb \delta$, known approaches and properties for stationary time series can be applied.  In particular, the large sample distribution of the discrete Fourier transform of a stationary time series \citep[Theorem 4.4.1]{brillinger2001} that provides the Whittle likelihood \citep{whittle1953} allows for the formulation of a product of Whittle likelihoods.  Define the local discrete Fourier transform at frequency $\ell$  within segment $q$ as
\begin{equation*}
\mb y_{q \ell} =  n_{q}^{-1/2} \sum_{t = \delta_{q-1} + 1}^{\delta_{q}} \mb X_{t} \exp(-2 \pi i  \omega_{q \ell} t), \quad \ell=1, \dots, L_q, \, q=1, \dots, m,
\end{equation*}
where $n_{q} = \delta_q - \delta_{q-1}$ is the number of observations in segment $q$, $\omega_{q \ell}=\ell/n_{q}$ are the Fourier frequencies, and $L_q = \lfloor(n_{q}-1)/2\rfloor$.  Given a partitioning into $m$ segments $\mb \delta$, $\mb y_{q \ell}$ are approximately independent zero-mean complex multivariate Gaussian random variables with covariance matrices $f(u_q, \omega_{q\ell})$, leading to a log-likelihood that can be approximated by a sum of log Whittle likelihoods
\begin{equation*}\label{likeihood1}
{\cal L}( Y \mid f, {\mb \delta}, m) \approx - \sum_{q=1}^{m} \sum_{\ell=1}^{L_q} \left\{ \log  \left|f\left( u_q, \omega_{q \ell} \right) \right| + \mb y_{q \ell}^* f^{-1}\left(u_q, \omega_{q \ell} \right) \mb y_{q \ell} \right\}.
\end{equation*}
Throughout this article, we use the notation where $Y$ represents the collection of all local discrete Fourier transforms.

It should be noted that, as opposed to methods such as those considered by \cite{ombao2005} and \cite{davis2006}, where the partitions are treated as fixed, here the number of segments $m$ and partition $\mb \delta$ are random variables whose prior distributions are given in Section \ref{subsec:prior}.   Estimators and inference will be obtained by averaging over the posterior distribution of partitions and number of segments.


\subsection{The Modified Cholesky Decomposition}\label{subsec:cholesky}
The spectral matrix is positive definite and, to nonparametrically allow for flexible smoothing among the different components while preserving positive-definiteness, we model modified Cholesky components of local spectra via linear penalized splines.
The use of other possible nonparametric local spectrum estimators is discussed in Section \ref{remarks}.
The modified Cholesky decomposition represents a time-varying spectral matrix  as
\begin{equation*}
f^{-1}(u, \omega)=  \Theta(u,\omega) \Psi(u,\omega)^{-1} \Theta(u,\omega)^*
\end{equation*}
for a complex-valued $N \times N$ lower triangular matrix $\Theta(u, \omega)$ with ones on the diagonal and a positive diagonal matrix $\Psi(u, \omega)$.  For a piecewise stationary approximation with partition $\mb \delta$ into $m$ segments, we define the local modified Cholesky decomposition as $f^{-1}(u_q, \omega)=  \Theta(u_q,\omega) \Psi(u_q,\omega)^{-1} \Theta(u_q,\omega)^*$, for $q=1, \dots, m$, and let $\theta_{jk q}(\omega)$ and $\psi_{jjq}(\omega)$ be the $jk$ and $jj$ elements of $\Theta(u_q,\omega)$ and $\Psi(u_q,\omega)$, respectively.   Then, for each segment, there are $N^2$--components to estimate:  $\Re\{\theta_{jk q}(\omega)\}$ for $j > k =1, \dots, N$, $\Im\{\theta_{jk q}(\omega)\}$ for $j > k =1, \dots, N$, and $\psi_{jjq}(\omega)$ for $j=1, \dots, N$.

In addition to the restriction that spectral matrices are positive definite, they also possess restrictions as functions of frequency in that they are periodic and  are Hermitian.  This means that all components are periodic, real components of $\Theta$ and of the diagonal $\Psi$ are even functions, or $\Re\{\theta_{jk q}(\omega)\} = \Re\{\theta_{jk q}(-\omega)\}$ and $\psi_{jjq}(\omega) = \psi_{jjq}(-\omega)$, and imaginary components are odd, or  $\Im\{\theta_{jk q}(\omega)\} = -\Im\{\theta_{jk q}(-\omega)\}$.  We model the components using periodic even and odd linear splines by considering
\begin{eqnarray}
\label{one}
\Re\{\theta_{jk q}(\omega)\}\ &=& c_{jk q 0} + \sum_{s=1}^{S-1} c_{jk q s} \cos(2\pi s \omega), \\
\label{two}
\Im\{\theta_{jk q}(\omega)\}\ &=& \sum_{s=1}^{S} b_{jk q s} \sin(2\pi s \omega),\\
\label{three}
\log\{\psi_{jj q}(\omega)\} &=& d_{jj q 0} + \sum_{s=1}^{S-1} d_{jjqs} \cos(2\pi s \omega).
\end{eqnarray}
The Fourier frequencies for each segment form an equally spaced grid, so that the Demmler-Reinsch bases for periodic even and odd smoothing splines for local periodograms are given by $\left\{\cos(2\pi s \omega);\, s=0,\dots,(L_q-1)\right\}$ and $\left\{\sin(2\pi s \omega) ; s=1,\dots, L_q\right\}$, respectively \citep[Section 3]{schwarz2016}.  Since $\psi_{jj}(\omega)$ is positive, it is modeled on the log-scale.

Not all Cholesky components are required to change at each partition point, enabling components to evolve differently over time.  In Section \ref{subsec:prior}, we define prior distributions for the components that change at each partition point, in addition to priors for the number and location of partition points and for the coefficients $b_{jk q s}$, $c_{jk q s}$ and $d_{j j q s}$.   Prior distributions for the coefficients are selected to regularize integrated squared first derivatives and formulate Bayesian linear penalized splines.  It should be noted that this  Bayesian linear penalized spline model for local spectra differs somewhat from that used by \citet{rosen2007} for stationary time series.
They use a model that is periodic, but not restricted to be odd or even.   Accounting for these geometric restrictions has been shown to improve performance, especially at the boundary \citep{krafty2013}.
The proposed model for local spectra is equivalent to the smoothing spline model for a stationary spectrum of \citet{krafty2013} when $S=L_q$.  Since coefficients decay rapidly, using $S < L_q$ basis functions presents considerable computational savings without sacrificing model fit.  In simulation studies and data analysis, we select $S=10$, which accounts for at least 99.975\% of the total variance of the full smoothing spline when $T \le 10^4$ \citep{krafty2016}.

\section{Priors and Sampling Scheme}\label{sec:prior}
\subsection{Priors}\label{subsec:prior}
We first define prior distributions for the parameters of the temporal partitioning, then define priors for the parameters of the penalized spline model conditional on the partitioning.  The time series is partitioned into $m$ approximately stationary segments and we choose a discrete uniform $\mathcal{U}(1,M)$ prior for $m$, where $M$ is a fixed large integer that represents a maximum number of possible segments.  To assure that the local Whittle likelihood approximation holds, we chose a minimum number of time points per segment $n_{\min}$ and select $M \le \lfloor T/n_{\min} \rfloor$.
Conditional on the number of segments $m$, the prior for the partition $\mb \delta$ is
\begin{equation*}
\pr(\mb \delta \mid m) = \prod_{q=1}^{m-1} \pr(\delta_{q} \mid \delta_{q-1}, m),
\end{equation*}
where $\pr(\delta_{q} \mid \delta_{q-1}, m) = 1/\alpha_{q}$ and $\alpha_{q}=T-\delta_{q-1}-(m-q+1)n_{\text{min}}+1$ is the number of possible locations for the $q$th partition point.  This prior places equal weight on all possible locations of a partition point conditional on previous partition points.
Some Cholesky components can remain unchanged across successive segments to enable flexible smoothness across time and we assume a uniform prior for the set of components that change at a given partition point.  More specifically, at least one of the $N^2$--Cholesky components is required to change across a partition point, so that there are a total of $2^{N^2}-1$ possible sets of Cholesky components that can change.  Letting $\phi_{q}$ be the set of $N^2$--Cholesky components that change at $\delta_q$, we assume a prior distribution on $\phi_{q}$ such that probability of any nonempty set of the $N^2$--components is $1/(2^{N^2} - 1)$.

The Cholesky components of local spectra within segments are modeled using the Demmler-Reinsch basis for linear splines so that smoothing priors that regularize the integrated square first derivative can be formulated as Gaussian priors with diagonal covariance matrices \citep{rosen2007}. Let $ \mb c_{jkq}=\{ c_{j k q 0}, \dots, c_{jkq(S-1)}  \}'$, $\mb b_{jkq}=\{ b_{jkq1}, \dots, b_{jkqS}  \}'$, and  $\mb d_{jjq}=\{ d_{jjq0}, \dots, d_{jkq(S-1)}  \}'$ be vectors of coefficients of basis functions for the Cholesky components within the $q$th segment defined in (1), (2), and (3). The priors on $\mb c_{jkq}$, $\mb b_{jkq}$, and $\mb d_{jjq}$ are $N(0,D_{cjkq})$, $N(0,D_{bjkq})$ and $N(0,D_{djjq})$, respectively, where
\begin{eqnarray*}
	D_{cjkq}&=&\diag{[\sigma_{\alpha}^2, \lambda^2_{cjkq}(2 \pi 1 )^{-2}, \dots, \lambda^2_{cjkq} \{2 \pi (S-1)\}^{-2}]}, \\
	D_{bjkq}&=&\diag{\{\lambda^2_{bjkq}(2 \pi 1 )^{-2}, \dots, \lambda^2_{bjkq}(2 \pi S)^{-2}\}}, \\
	D_{djjq}&=&\diag{[\sigma_{\alpha}^2, \lambda^2_{djjq}(2 \pi  1)^{-2}, \dots, \lambda^2_{djjq}\{2 \pi (S-1)\}^{-2}}].
\end{eqnarray*}
Condition on smoothing parameters $\lambda_{cjkq}^{2}$, $\lambda_{bjkq}^{2}$ and $\lambda_{djjq}^{2}$, these prior distributions induce priors on the integrated square first derivatives of local Cholesky components that are proportional chi-square distributions where the smoothing parameter is the proportionality constant.  Consequently, a smoothing parameter controls the roughness of a local Cholesky component such that, as its smoothing parameter is shrunk towards zero, the component tends towards a constant function of frequency with probability 1.  We assume priors for $\lambda_{cjkq}^{2}$, $\lambda_{bjkq}^{2}$ and $\lambda_{djjq}^{2}$ to be $\text{Uniform}(0,\kappa)$, where $\kappa$ is a known large constant. The parameter $\sigma_{\alpha}^2$ is a hyperparameter for the prior variance of the intercepts of the Cholesky components and is set to be a fixed large value.   We found results to be insensitive to the selection of hyperparaemters, with  $\kappa \in (10^3, 10^6)$ and  $\sigma_{\alpha}^2 \in (10^3, 10^7)$ providing indistinguishable results in all empirical examples considered.

\subsection{Description of the Sampling Scheme}\label{subsec:sampling}

We present a  scheme  based on RJMCMC techniques to sample from the joint posterior distribution of parameters \citep{green1995}.  Each iteration consists of two types of moves: within-model and between-model moves.  We briefly describe the sampling scheme in this section and provide technical details in the supplementary materials.

Between-model moves involve randomly proposing either a death where the number of segments $m$ decrease by 1, or a birth where $m$ increases by 1. Parameters drawn within the move are jointly accepted or rejected. We denote $m^p$ and $m^c$ as the proposed and current number of segments respectively.  If a birth move is proposed, $m^p=m^c+1$, and a new partition point is drawn by first sampling a current segment to be split, then randomly selecting the new partition point within the segment. Conditional on $m^p$ and the new partition point, the set of Cholesky components that change at the new partition point, $\phi_{q}$, is sampled from the discrete uniform distribution. Then, in each of the two new segments, the smoothing parameters corresponding to the Cholesky components that were selected to change between the segments are proposed. Finally, coefficients of basis functions of the selected Cholesky components for both new segments are drawn from approximated conditional Gaussian distributions.

If a death is proposed, then $m^p=m^c-1$. A partition is randomly selected to be deleted and a new segment is formed by combing the two neighboring segments separated by this point. Smoothing parameters for components that were previously different across these two segments are used to form the corresponding smoothing parameters in the newly combined segment. Then, the corresponding coefficients of basis functions of the combined segments are drawn.

Within-model moves involve no change in the number of partitions or segments, namely, $m^p=m^c$.  First, a partition $\delta_{q}$ is randomly selected and relocated. Then, basis function coefficients of all Cholesky components within the segments impacted by the move are drawn.
To efficiently explore local parameter spaces,  Hamiltonian Monte Carlo (HMC) \citep{neal2011, gelman2013} is used to draw coefficients of basis functions of the Cholesky components.  Lastly, impacted smoothing parameters are drawn from conditional inverse gamma distributions.

It is worth noting that our proposed method has a number of differences compared to the adaptive Bayesian method proposed by \cite{zhang2016}.   First, the proposed model and sampling scheme allow some components to remain stationary across a partition point through the parameter $\phi$.  This allows estimated spectral components to have different temporal structures.  In a manner similar to other existing methods \citep{ombao2001, davis2006}, the method of \cite{zhang2016} requires all components to change at each partition point.  Second, the unknown number and location of partitions are sampled via two different methods, RJMCMC and SSAMC, respectively. While SSAMC has been shown to improve upon the convergence of RJMCMC in some circumstances, particularly in highly complex sample spaces,  empirical results suggest that this is not necessarily the case in many examples of nonstationary multivariate spectrum analysis.
\cite{liang2009} noted that the smoothing steps in SSAMC could potentially hinder estimation when there exist abrupt jumps between different models, which is the case when a time series is piecewise stationary with abrupt changes. In this setting, one model is of most interest. Thus, once the sampler arrives at the true number of segments and roughly correct partition locations, ideally the sampler will stay at this ``correct model" with a high probability rather than frequently jumping to other models. On the other hand, when we have a time series with a slowly-varying spectrum, SSAMC could be superior to RJMCMC in terms of moving between different models. However, within--model moves are more crucial in obtaining estimates that can, in a frequentist sense, effectively approximate slowly--varying patterns, as one desires the partitions to have a high probability of relocating.
Lastly, in the proposed method, within-model moves are sampled via HMC.  The HMC techniques enable desirable within-model convergence and acceptance rates, and consequently enable the location of partition points to vary with sufficient probability for final estimates averaged over all iterations to effectively approximate slowly--varying changes.

\section{Simulation Studies}\label{sec:example}

In this section, we perform simulation studies under three settings to investigate the frequentist characteristics of the proposed method and to compare it to existing methods.  The goal of the first two sets of simulations is to investigate performance in estimating time-vary spectra, first  for  a trivariate piecewise stationary process with one partition and then for a bivariate slow-varying process.   The goal of the third setting is to investigate properties in identifying partition points for a piecewise stationary process with multiple partitions of varying magnitudes.
For all models, results are reported where hyperparameters of the proposed method were chosen as $n_{\min} = 60$, $\sigma^2_{\alpha}=10^4$ and $\kappa=10^5$.

\subsection{Piecewise Stationary Process}\label{subsec:var2}
\begin{figure}
	\centering
	\includegraphics[width=4in]{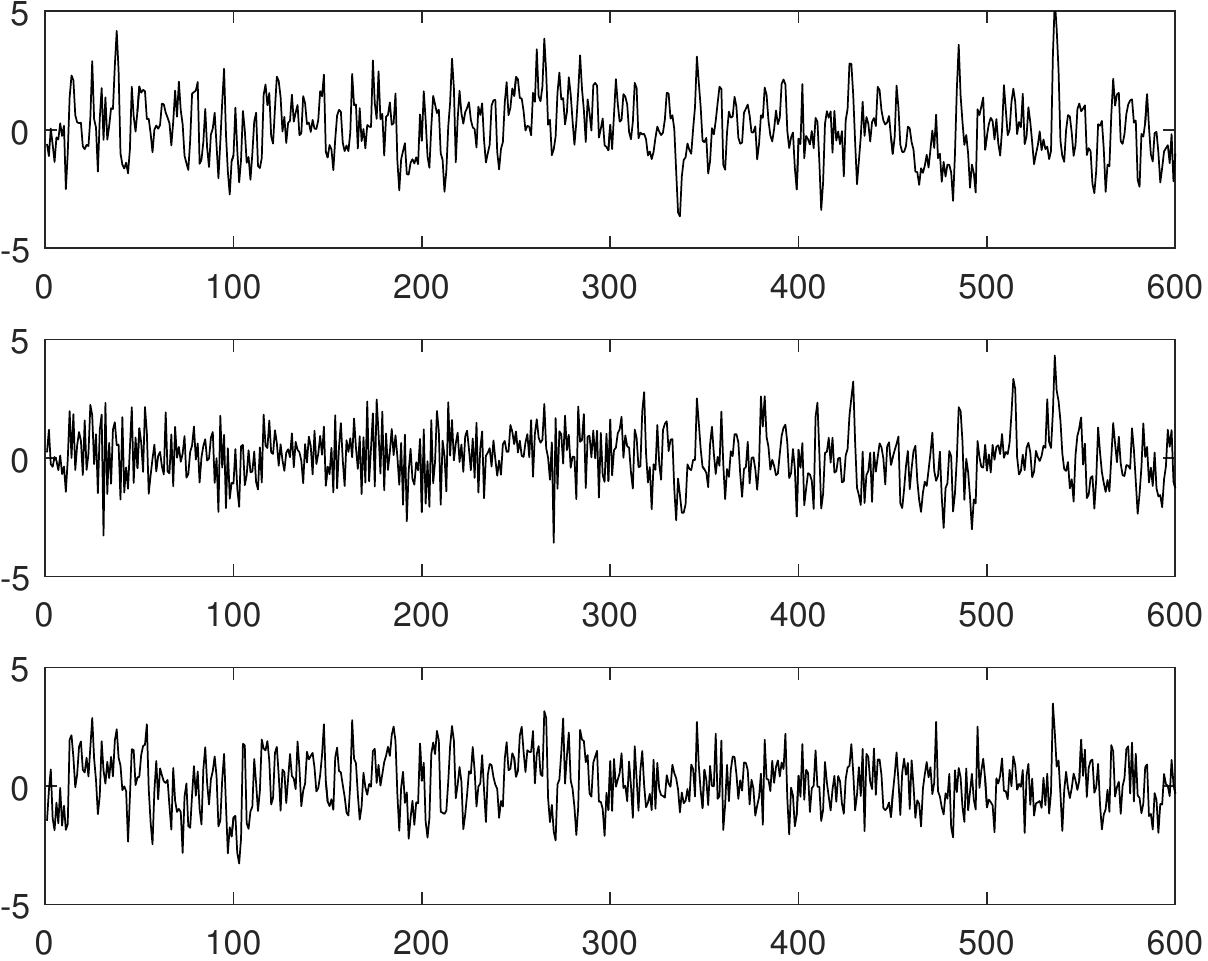}
	\caption{A simulated piecewise stationary trivariate time series.}
	\label{simtimevar}
\end{figure}

We simulated 250 independent trivariate piecewise stationary time series of length $T=600$ as
\begin{equation*}\label{eq:simvar}
X_t =
\begin{cases}
\mb \epsilon_{t} + \Phi_{11} \mb \epsilon_{t-1} + \Phi_{12} \mb\epsilon_{t-2}   & \quad \text{if }  1 \le t \le 300\\
\mb \epsilon_{t} + \Phi_{21} \mb \epsilon_{t-1} + \Phi_{22} \mb \epsilon_{t-2}   & \quad \text{if } 301 \le t \le 600,
\end{cases}
\end{equation*}
where
\begin{equation*}
 \Phi_{11} = \begin{pmatrix}
0.6 & 0  &  0 \\
0.2 & -0.5  &  0  \\
0.1 & 0.3 & 0.4 \\
\end{pmatrix},
\Phi_{21} = \begin{pmatrix}
0.6 & 0  &  0 \\
0.2 & 0.5  &  0  \\
-0.1 & -0.3 & 0.4 \\
\end{pmatrix},
\end{equation*}
$\Phi_{12} = \Phi_{22} = \diag(0.3, 0.3, 0)$, and $\mb \epsilon_{t}$ are independent zero-mean trivariate Gaussian random variable whose components have unit variance and pairwise correlation $0.5$.  The true power spectrum is  $f(u,\omega) = \Phi(u,\omega) \Sigma \Phi(u, \omega)^*$ where  $\Phi(u,\omega) = I + \Phi_{11} \exp(-2 \pi i \omega) + \Phi_{12} \exp(-4 \pi i \omega)$ for $u \in [0, 1/2]$, $\Phi(u,\omega) = I + \Phi_{21} \exp(-2 \pi i \omega) + \Phi_{22} \exp(-4 \pi i \omega)$ for $u \in (1/2, 1],$ and $\Sigma = \cov(\mb \epsilon_{t})$ \citep[Chapter 9.4]{priestley1981}. A simulated realization of this process is shown in Figure \ref{simtimevar}.

\begin{figure}
	\centering
	\begin{minipage}[b]{.82\textwidth}
		\includegraphics[width=\textwidth]{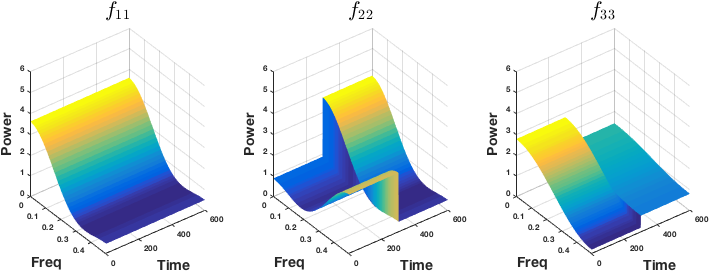}
	\end{minipage}

		\begin{minipage}[b]{.82\textwidth}
			\includegraphics[width=\textwidth]{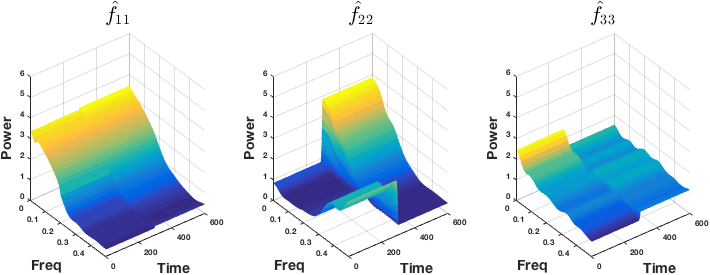}
		\end{minipage}
	
	\begin{minipage}[b]{.82\textwidth}
		\includegraphics[width=\textwidth]{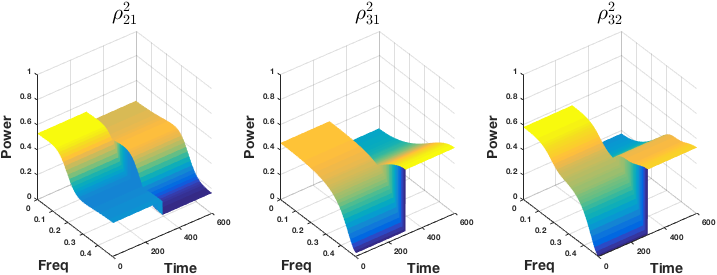}
	\end{minipage}

		\begin{minipage}[b]{.82\textwidth}
			\includegraphics[width=\textwidth]{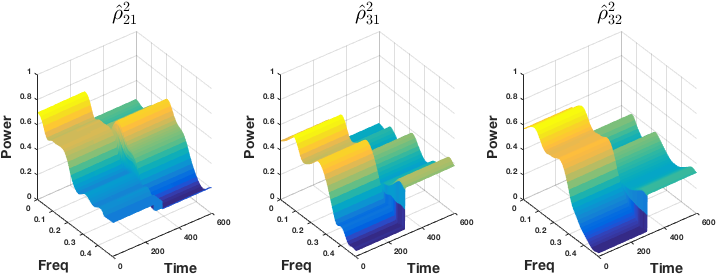}
		\end{minipage}
	\caption{Time-varying individual spectra and their estimates (top two rows), and time-varying coherences and their estimates (bottom two rows) from the piecewise stationary process in Figure \ref{simtimevar}. }
	\label{spectAR}
\end{figure}

We investigate the performance of an estimator of a spectral matrix $f(u,\omega)$ through estimates of individual spectra, $f_{11}(u,\omega)$, $f_{22}(u,\omega)$, and $f_{33}(u,\omega)$, and of the coherences
\[\rho^2_{j k}(u, \omega) = \left|f_{j k}(u, \omega)\right|^2/\left\{f_{jj}(u, \omega) f_{kk}(u, \omega)\right\},~~~  \text{for} ~~ j = 2,3 ~~ \text{and}~~ \forall k < j.\]
The individual spectra and coherences for the piecewise process and estimates from the realization displayed in Figure \ref{simtimevar} using the proposed method are given in Figure \ref{spectAR}.

For each simulated time series, the time-varying power spectrum was estimated by three methods: the proposed method with $12000$ iterations and a burn-in period of $4000$ iterations, the piecewise vector autoregressive method of \citet{davis2006} with default tuning parameters, and the nonparametric smoothing spline ANOVA approach of \citet{guo2006} with 10 prespecified blocks.  Relative performance of the estimators were assessed through average square error (ASE) averaged across the equally spaced grid of $T$ time points and 51 frequency values. The means and standard deviations of the ASE are presented in Table~\ref{simtab}.   It should be noted that the SSAMC-based method of \cite{zhang2016}, whose empirical performance is investigated in Section \ref{sec:multimodal} for a smaller number of replicates, was not implemented in this setting due its high computational cost, which makes its implementation impractical for 250 replicates.

Unsurprisingly, the smoothing spline ANOVA estimator, which is based on a slowly-varying model, exhibited poor performance relative to the other estimators in this abruptly changing setting.  This is particularly true when estimating $f_{2 2}$, $\rho^2_{31}$ and $\rho^2_{32}$, which experience abrupt and drastic changes.  The proposed method had smaller mean ASE compared to the piecewise vector autoregressive method for all components.
This can be attributed to the fact that, while the piecewise vector autoregressive estimator requires all components to change at each partition point, the proposed approach enables some components to remain unchanged.  This allows the proposed method to more accurately estimated components that remains constant across time and functions thereof.

\begin{table}[t]
	\centering
	\caption { \label{simtab} Results of the simulation study of the piecewise stationary process. Based on 250 repetitions, means (standard deviation) of average square error $\times10^{2}$ of spectral estimates obtained through the proposed method,  piecewise vector autoregressive modeling (Auto-PARM), and smoothing spline ANOVA (SmoothANOVA).}{%
		\begin{tabular}{l c c c c c c c c c c c}
			&  $f_{11}$ & $f_{22}$  & $f_{33}$& $\rho^2_{2 1}$ & $\rho^2_{3 1}$ & $\rho^2_{3 2}$ \\ \hline
			Proposed & 25.1 (28.6)    & 22.8 (20.5)   &   13.8  (13.3) & 1.4 (0.8) & 2.1 (0.9) & 2.2 (1.1)\\
			Auto-PARM   	& 37.2 (33.2)    & 25.4 (16.2)  & 14.3 (7.7)  & 6.2 (2.4)   & 3.8 (1.1) & 9.0 (1.3)\\
			SmoothANOVA     & 39.8 (10.6)    & 93.4 (16.2) & 27.3 (6.4)  & 9.6 (0.5) &12.8 (0.9) & 13.4 (1.2)\\ \hline
		\end{tabular}
	}\end{table}	

In addition,  percentiles of the empirical distribution of the simulated sample provide a natural means to conduct inference on any function of spectral matrices.  To evaluate frequentist performance in conducting inference, we computed  95\% pointwise credible intervals for spectrum components and evaluated average coverage across time-frequency points.  The mean (standard deviation) coverage of the 95\% credible intervals for $f_{11}$,  $f_{22}$, $f_{33}$, $\rho^2_{21}$, $\rho^2_{31}$, and $\rho^2_{32}$ were 95.2\% (2.8\%), 94.1\% (3.4\%), 95.8\% (2.7\%), 93.5\% (3.2\%), 92.7\% (2.4\%), and 92.4\% (2.3\%), respectively.

\subsection{Slow-varying Process}
\begin{figure}
	\centering
	\includegraphics[width=4in]{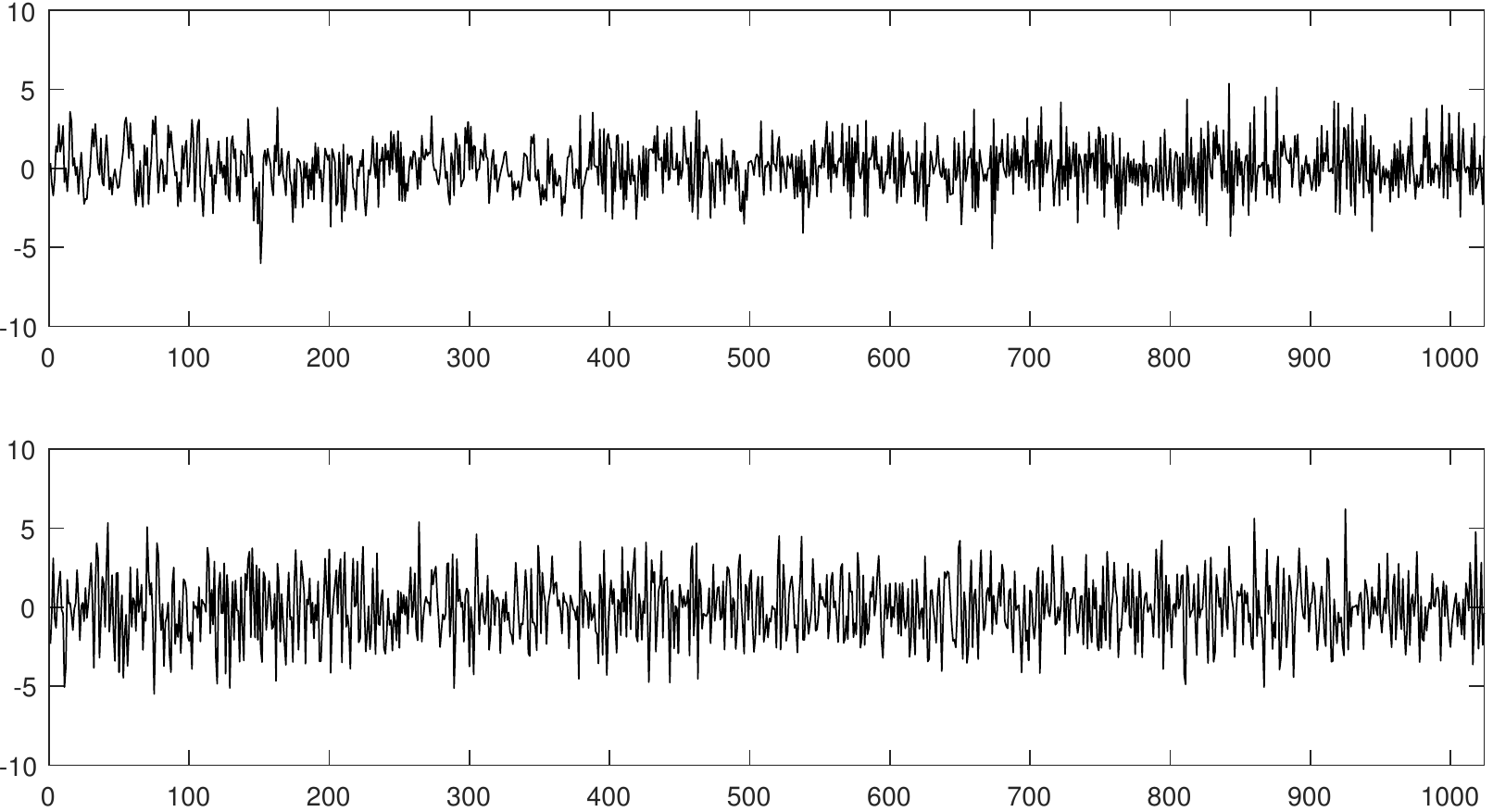}
	\caption{A simulated slow-varying bivariate time series.}
	\label{simslowma}
\end{figure}

\begin{figure}[t]
	\centering
	\begin{minipage}[b]{.8\textwidth}
		\includegraphics[width=\textwidth]{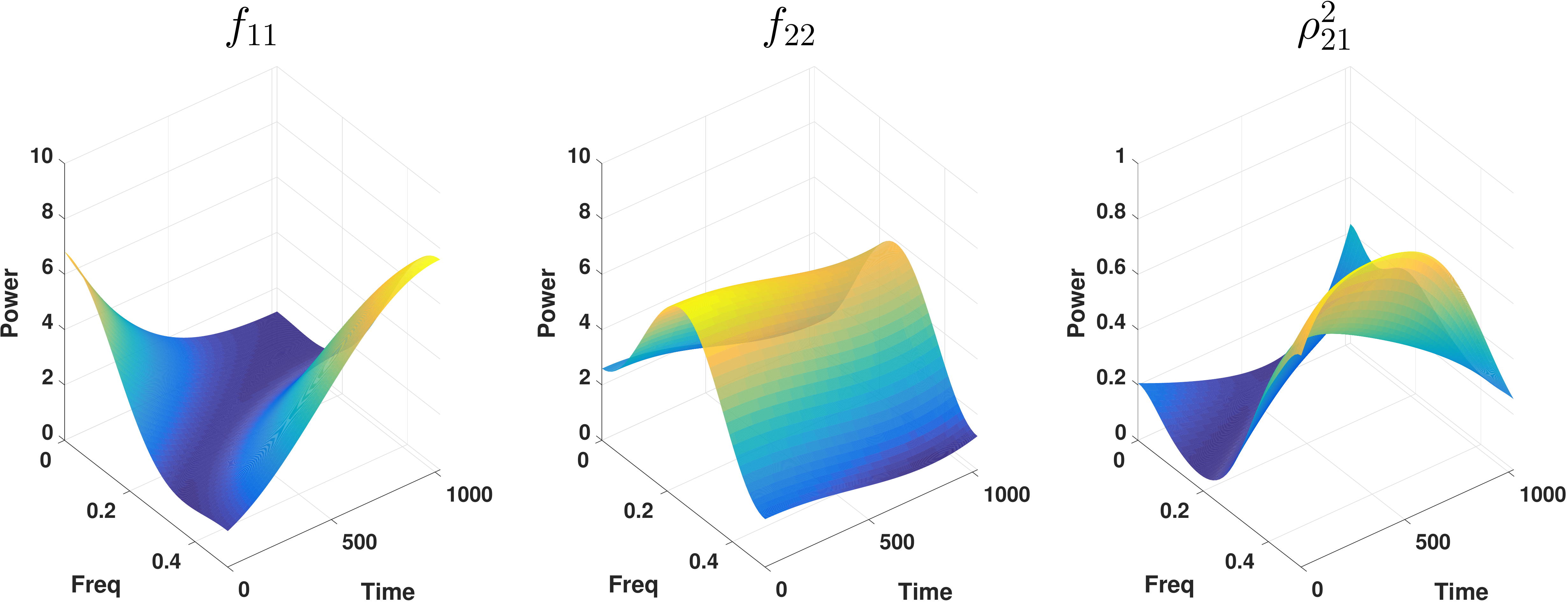}
	\end{minipage}
	
	\begin{minipage}[b]{.8\textwidth}
		\includegraphics[width=\textwidth]{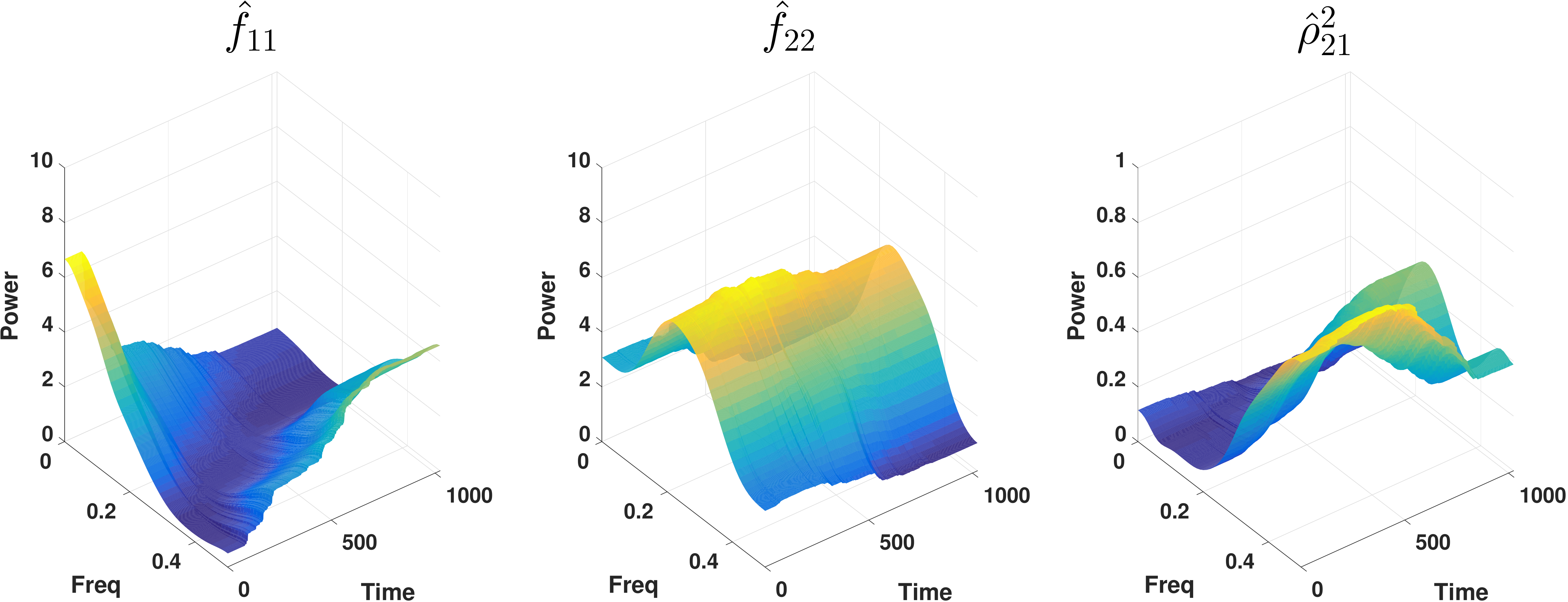}
	\end{minipage}
	\caption{Time-varying individual spectra and coherence (top row), and estimates (bottom row) from the slowly-varying process time series in Figure \ref{simslowma}. }
	\label{spectslowma}
\end{figure}

In this section, we considered a bivariate process that possesses a slow-varying power spectrum. We simulated 250 time series as
\begin{equation*}\label{eq:simslowma}
X_t = \mb \epsilon_t + \Phi_{1 t} \mb \epsilon_{t-1} + \Phi_{2 t} \mb \epsilon_{t-2}, \quad t=1, \dots, 1024,
\end{equation*}
with
$$
\Phi_{1 t} = \begin{pmatrix}
\phi_1(t) &~ -1     \\
-1 &~ \phi_{2}(t)      \\
\end{pmatrix},
\Phi_{2 t} = \begin{pmatrix}
0.5 &~ 0      \\
0 &~ -1.2      \\
\end{pmatrix},
$$
where $\phi_{1}(t)=1.122 \left \{ 1-1.781 \sin(\pi t/2048)  \right \}$, $\phi_{2}(t)=1.122 \left \{ 1-1.781 \cos(0.8\pi t/2048)  \right \}$, and $\mb \epsilon_t$ are independent zero-mean bivariate Gaussian random variables whose components have unit variance and pairwise correlation $0.2$.  A realization of the time series is plotted in Figure  \ref{simslowma}; the true time-varying individual spectra, coherence and estimates under the proposed procedure from this realization are displayed in Figure \ref{spectslowma}.
As in Section \ref{subsec:var2}, three methods were compared: the proposed method with $10000$ iterations and a burn-in period of $2000$ iterations, the piecewise vector autoregressive method of \citet{davis2006} with default tuning parameters, and the nonparametric smoothing spline ANOVA approach of \citet{guo2006} with 16 predefined blocks. Table \ref{simtab2} displays the means and standard deviations of the ASE for each of the three estimators. Both the proposed method and the smoothing spline ANOVA method outperformed the piecewise vector autoregressive approach in all components.  Although the proposed procedure and the piecewise autoregressive estimator are both based on piecewise stationary approximations, the proposed method adaptively averages over the distribution of partitions to better capture the slowly-varying structure.  The proposed method, which uses the Whittle likelihood, performed  slightly better than the smoothing spline ANOVA approach, which utilizes penalized sums of squares.  This is consistent with findings for stationary univariate \citep{pawitan1994}, nonstationary univariate \citep{qin2009wang} and stationary multivariate \citep{krafty2013} time series, in which Whittle likelihood based methods were found to be more efficient compared to sums of squares based methods. The mean (standard deviation) of the average coverage of 95\% pointwise credible intervals for $f_{11}$,  $f_{22}$, and $\rho^2_{2 1}$ under the proposed method are 96.7\% (2.3\%), 96.5\% (2.1\%), 94.8\% (4.6\%), respectively.

\begin{table}[t!]
	\centering
	\caption{ \label{simtab2} Results of the simulation study of the slowly-varying process. Based on 250 repetitions, means (standard deviation) of average square error $\times10^{2}$ of spectral estimates obtained through the proposed method,  piecewise vector autoregressive modeling (Auto-PARM), and smoothing spline ANOVA (SmoothANOVA).}
{%
		\begin{tabular}{l c c c c c c c}
			&  $f_{11}$ & $f_{22}$  & $\rho^2_{2 1}$ \\ \hline
			Proposed & 49.2 (19.3)    & 51.4 (26.3)      & 1.1  (0.2)  \\
			Auto-PARM   	& 90.6 (26.6)    & 85.5 (34.5)    & 7.0 (3.2)   \\
			SmoothANOVA     & 51.8 (20.3)    & 61.2 (22.5)     & 1.6 (1.3)   \\ \hline
		\end{tabular}

	}\end{table}

\subsection{Estimating Multiple Partitions}\label{sec:multimodal}

In this section, the main focus is to investigate the frequentist properties of the proposed method in estimating the number and location of the partition points, and to compare them to the SSAMC-based method of \cite{zhang2016}. We simulate 20 time series from the following bivaraite piecewise autoregressive process
\begin{equation}\label{eq:multimodal}
X_t = \Phi_{1} X_{t-1} + \Phi_{2} X_{t-2} + \mb \epsilon_{t},
\end{equation}
where
\begin{equation*}
\begin{cases}
\Phi_{1}= \diag(0.5, -0.6),~ \Phi_{2} = \diag(0,-0.5),~ \mb \epsilon_{t} = \mb \epsilon_{t}^1  & \quad \text{if }  1 \le t \le 400\\
\Phi_{1}= \diag(0.5, 0.6),~ \Phi_{2} = \diag(0,-0.5),~ \mb \epsilon_{t} = \mb \epsilon_{t}^1 & \quad \text{if } 401 \le t \le 5000 \\
\Phi_{1}= \diag(0.5, 0.6),~ \Phi_{2} = \diag(0,-0.5),~ \mb \epsilon_{t} = \mb \epsilon_{t}^2  & \quad \text{if } 5001 \le t \le 10000\\
\Phi_{1}= \diag(1.32, 0.6),~ \Phi_{2} = \diag(-0.81,-0.5),~ \mb \epsilon_{t} = \mb \epsilon_{t}^2  & \quad \text{if } 10001 \le t \le 12000,
\end{cases}
\end{equation*}
and $\mb \epsilon_{t}^1$, $\mb \epsilon_{t}^2$ are independent zero-mean bivariate Gaussian random variables whose components have unit variance and pairwise correlation $0.5$ and $0.8$, respectively. The partitions are located far apart from each other, and the change at $t=5000$, where there is only a slight change in off-diagonal elements of the error term, is subtle compared to the other changes.  

Both methods were run using $S=10$ basis functions for 10,000 iterations with a burn--in of 2,000. SSAMC is able to drawn multiple samples in a single iteration and we consider three cases for SSAMC: drawing 1 sample, drawing 5 samples, and drawing 20 samples. Other parameters are selected as the default suggested by \cite{zhang2016}.  Table  \ref{multi} reports the mean and standard deviation of estimated posterior probabilities of numbers of segments. The proposed method had highest estimated posterior probability of the correct number of segments with  $\widehat{\pr}(m=4|\mb X)=.794$.   The SSAMC sampler regularly transitioned between $m=4, 5, 6$ segments, with $\widehat{\pr}(m=4|\mb X)$ increasing with the number of samples. This finding coincides with the findings of \cite{liang2009}, in which SSAMC has a higher probability of sampling from different models compared to RJMCMC. The proposed estimator had smaller ASE compared to SSAMC, which could possibly be attributed to the proposed method sampling more often from models with the true number of $m=4$ segments compared to SSAMC. As more samples are drawn in each iteration, the ASE of SSAMC decreases.  Table \ref{multi} also reports the mean and standard deviation of run times (second) per iteration using R Version 3.4 and Windows 10 on a desktop computer with a 3.6 GHz Intel Core i7 processor and 8 GB RAM. The proposed method was observed to be considerably more efficient in terms of computation time.


%

\begin{table}[t]
	\centering
		\caption{ \label{multi} Results of the simulation study of process \eqref{eq:multimodal}  based on 20 repetitions: means (standard deviations) of posterior probabilities of numbers of segments $m$, of average square error (ASE) across all spectral components and of computational time per iteration for the proposed method and the SSAMC based method.}
		\vspace{.2cm}
	\scalebox{.8}{
\begin{tabular}{SSSSSSSS} \toprule
	{$m$} & {The propsed} & {SSAMC (1 sample)} &{SSAMC (5 Samples)} & {SSAMC (20 samples)} \\ \midrule
	1  & {0 (0)} & {0 (0)}  & {0 (0)} & {0 (0)}\\
	2  & {0 (0)} & {0 (0)}  & {0 (0)} & {0 (0)} \\
	 3  & {0 (0)}  & {0.001 (0.003)}  & {0 (0)}  & {0 (0)} \\
	  4  & {0.794 (0.325)}  & {0.251 (0.273)} &{0.306 (0.316)} & {0.385 (0.307)}    \\
	5  & {0.197 (0.334)}  & {0.348 (0.310)} & {0.342 (0.275)}  & {0.375 (0.248)}  \\
	 6  & {0.009 (0.017)}  & {0.275 (0.322)} &{0.288 (0.263)}  & {0.230 (0.290)}    \\
	 7  & {0  (0)} & {0.125 (0.172)} & {0.064 (0.108)}  & {0.010 (0.05)}    \\ \midrule
	   {ASE} & {1.665 (0.206)}  & {3.387 (0.461)} &  {2.934 (0.371)}  & {2.734 (0.329)}    \\
	     {Run Time (s)}  & {1.489 (0.217)}  & {5.119 (0.975)} & {22.337 (7.141)}  & {62.459 (22.066)}     \\ \bottomrule
\end{tabular}
}
\end{table}

\section{Applications}\label{sec:app}

\subsection{Analysis of EEG During Sleep}\label{sec:eeg}
Neurophysiological activity during sleep is hypothesized to be responsible for many of sleep's rejuvenating properties.  Consequently, to inform individualized treatments of disturbed sleep, physicians often use EEG to record patient's brain activity during sleep.  We consider data from one such patient at the Sleep Disorders Clinic at St Vincent's University Hospital, Dublin who was being treated for obstructive sleep apnea.

As one sleeps, the brain cycles between periods of rapid-eye movement sleep (REM), which is characterized by movement of the eyes and is when dreaming and body movement are more likely to occur, and non-rapid-eye movement sleep (NREM), which includes deep sleep.
Brain activity is inherently different during the different periods of sleep, so that the data are nonstationary; activity within each period and when transiting between periods plays a fundamental role in how sleep affects health and functioning.  In clinical practice,  EEG are visually inspected by trained technicians to identify periods of REM and NREM sleep \citep{iber2007}.  To gain a deeper understanding of brain activity during the transition from visually scored NREM to REM, we apply the proposed method to a 20 second long epoch of 2-channel (bivariate) EEG, with 10 seconds before and 10 seconds after the first visually scored transition between NREM into REM sleep. The EEG signal was recorded from bilateral central EEG leads with C3 and C4 referenced to linked mastoids A1 and A2 at a sampling rate of  128 HZ. The data were linearly detrended, filtered using a 0.02 HZ high-pass filter and a median filter for artifact removal, and standarized to unit variance.  The resulting data, which are displayed in Figure \ref{tseeg}, are a $N=2$ dimensional time series of length $T=2560$.

\begin{figure}
	\centering
	\begin{minipage}[b]{.9\textwidth}
		\includegraphics[width=\textwidth]{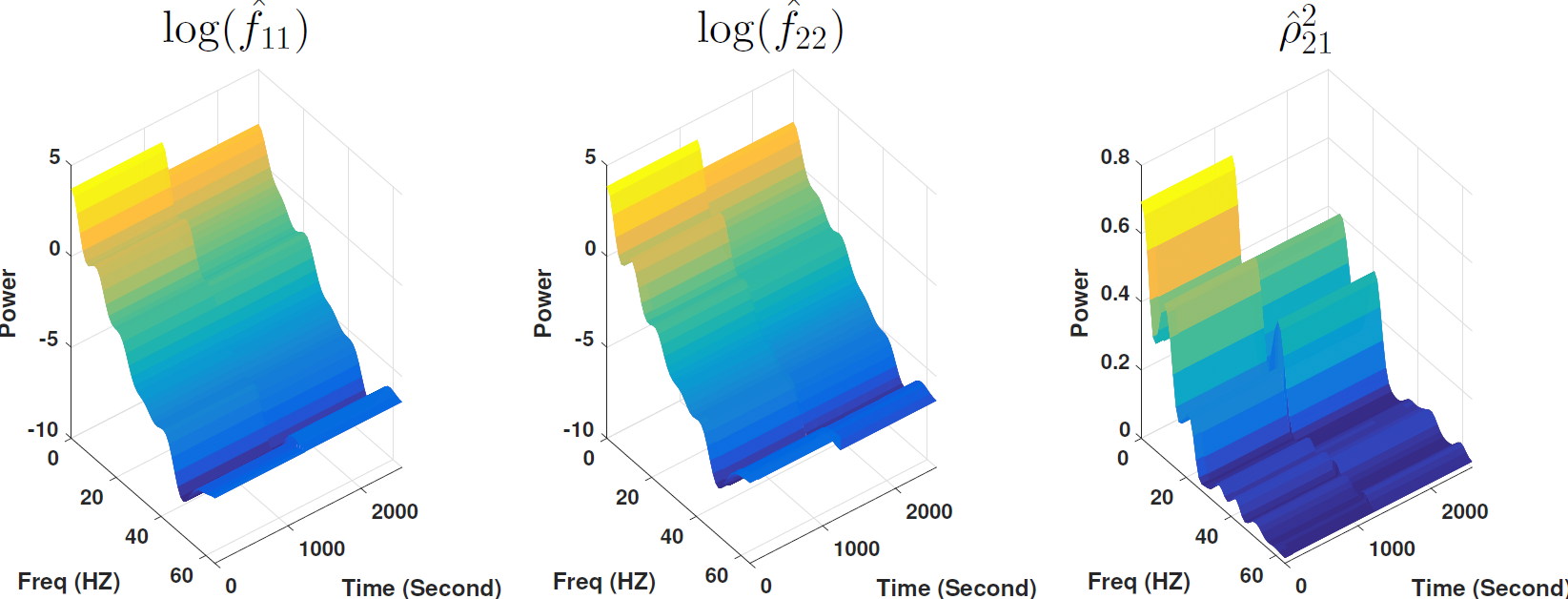}
	\end{minipage}
	\caption{Estimate of the time-varying log spectra and coherence of the EEG displayed in Figure \ref{tseeg} ($f_{11}$ and $f_{22}$ indicate C3--A2 and C4--A1 channels, respectively)}.
	\label{spectEEG}
\end{figure}

The proposed method was fit with hyperparameters $\sigma_{\alpha}^2=\kappa=10^5$, and run for 10,000 iterations with burin-in of 2,000.  The mean (standard deviation) for the run time per iteration under the computer specification reported in Section 4.3 was 0.463 (0.189) seconds.  Figure \ref{spectEEG} displays estimated log-spectra and coherence. Spectra are displayed on the log scale to aid visualization. There is an abrupt change both in the spectra and in the coherence. The change occurs at $t=1250$ (9.75 second) with a probability of 0.94, which is close to the visually scored transition from NREM sleep to REM sleep at $t=1280$ (10 seconds).  All components are nearly stationary before and after this change.  It should be noted that the proposed method provides an automated analysis of the nature of the temporal dynamics, providing evidence that the transition between sleep stages is sudden as opposed to gradually evolving.

Broadly, our results indicate that power at low frequencies decreases from NREM to REM sleep. The estimated log-spectra during NREM shows elevated power at 10--14 HZ.  Known as sleep spindles, activity within these frequencies are a common characteristic of deep sleep.   During REM, we observe characteristic peaks at beta power (16--31 HZ).  In addition to these characteristics of univariate spectra, the analysis also uncovered changes in coherence, where peak coherence shifts from the delta band of frequencies less than 4 Hz during NREM to the beta band of 16--31 HZ during REM. Further analyses of these data and convergence diagnostics are provided in supplementary material.
%
%

\subsection{Analysis of the El Ni\~no--Southern Oscillation}\label{sec:enso}

\begin{figure}[t]
	\centering
	\begin{minipage}[b]{.9\textwidth}
		\includegraphics[width=\textwidth]{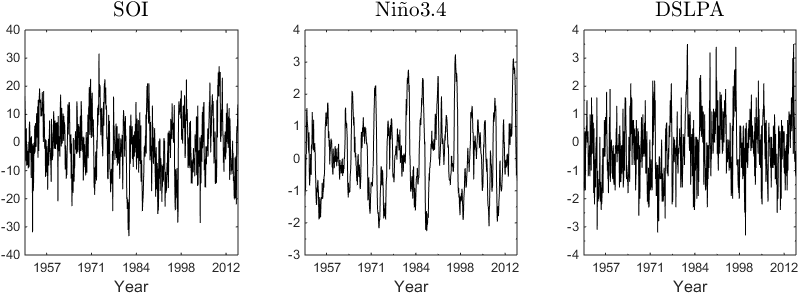}
	\end{minipage}
	
	\caption{Plots of montly SOI, Ni\~no3.4 and DSLPA from 1951 to 2016.}
	\label{SOI}
\end{figure}

The El Ni\~no--Southern Oscillation (ENSO) is an irregularly periodical disruption in the climate system over the Tropical Pacific, which has impacts on global weather and climate systems.   The process has not only be extensively studied in the geological literature, but also in the statistics literature
\citep{shumway2011, rosen2007, rosen2012}. However, these analyses have either been based on univariate time series analyses of individual ENSO indicators, or restricted by stationarity assumptions. In this section, we jointly consider three important indicative time series of ENSO observed from 1951 to 2016 (see Figure \ref{SOI}). The first indicator is the monthly Southern Oscillation Index (SOI), measuring mean sea-level pressure differences between Tahiti and Darwin station. The second indicator is Ni\~no3.4, the monthly average sea surface temperature in the region 5S-5N, 120W-170W. The third indicator is the monthly mean sea-level pressure anomalies at Darwin station (DSLPA).


\begin{figure}[t]
		\centering
		\begin{minipage}[b]{.55\textwidth}
			\includegraphics[width=\textwidth]{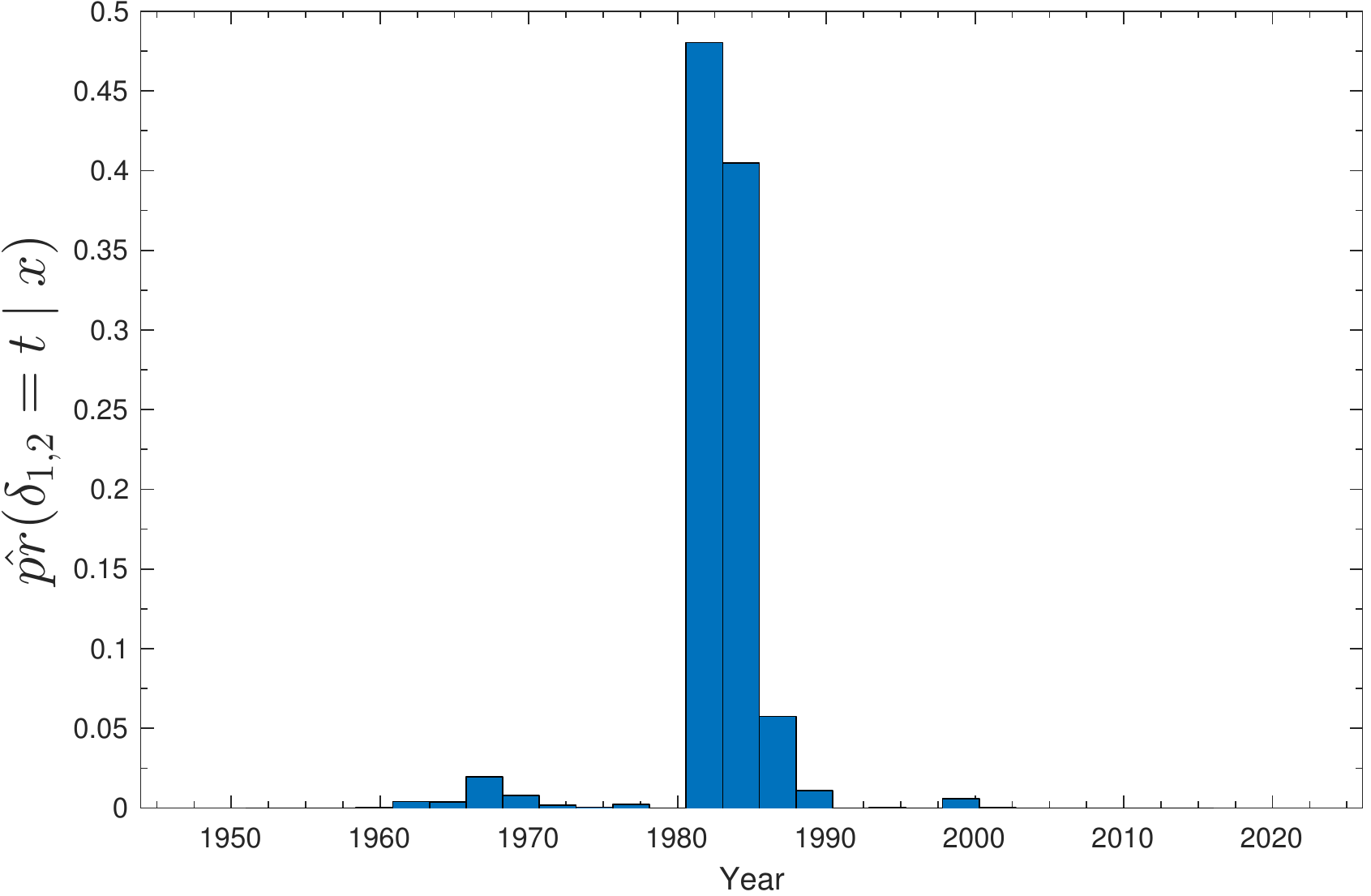}
		\end{minipage}
		\caption{Plot of $\hat{\pr}(\delta_{1,2}=t \mid x)$ for the ENSO analysis. }
		\label{hist}
	\end{figure}

We ran the sampling procedures for 12,000 iterations with burn-in of 4,000 iterations.  The mean (standard deviation) for the run time per iteration under the computer specification reported in Section 4.3 was 0.980 (0.505) seconds.
The estimated posterior probability of the number of segments is presented in Table \ref{segment}.  The mode is $m=2$ segments, which has an estimated posterior probability of .652, and the estimated posterior probability of $m=1$ segment, or stationarity, is .329.  The estimated the posterior distribution of the location of partitions conditional on $m=2$, $\hat{Pr}(\delta_{1,2}=t \mid x)$, is shown in Figure \ref{hist}.   The estimated conditional (on $m=2$) posterior probability of this point occurring between 1980--1989 is 0.940, with non-zero probability across the entire interval, but with higher probability closer to 1980. These results suggest that the spectrum could have experienced changes in  1980-1989.

\begin{figure}[t]
		\centering
		\begin{minipage}[b]{.9\textwidth}
			\includegraphics[width=\textwidth]{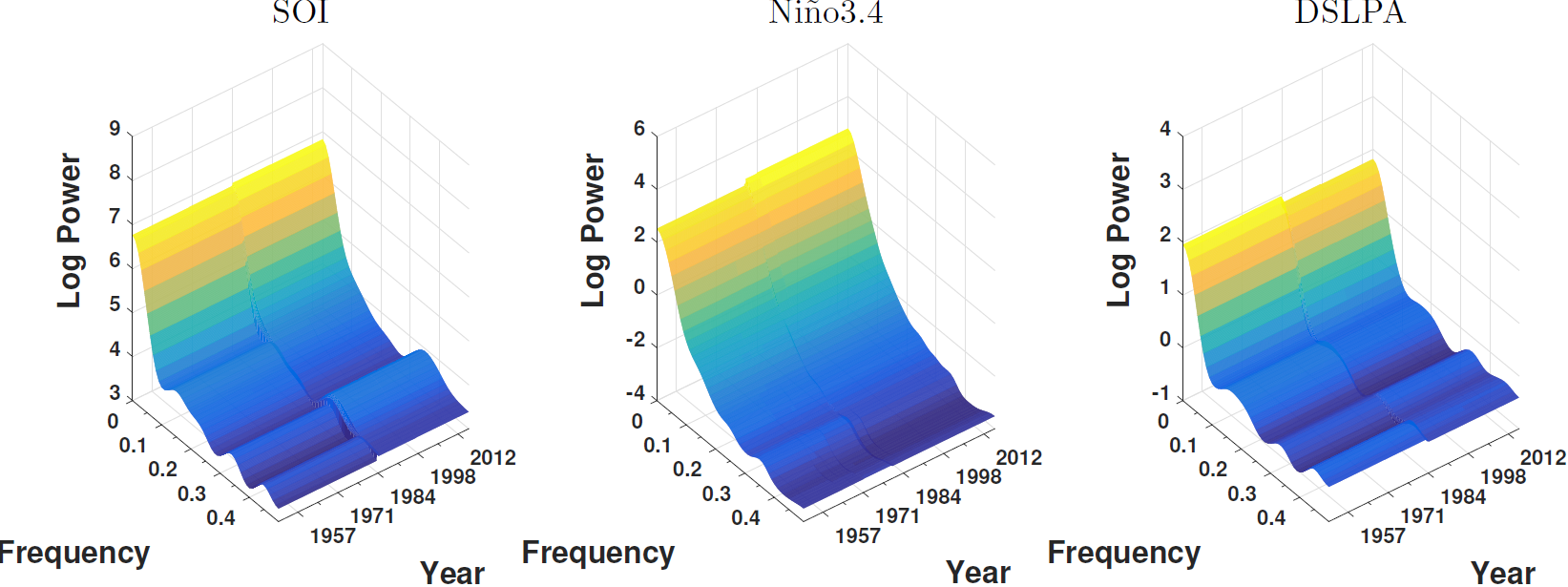}
		\end{minipage}
		
		\begin{minipage}[b]{.9\textwidth}
					\includegraphics[width=\textwidth]{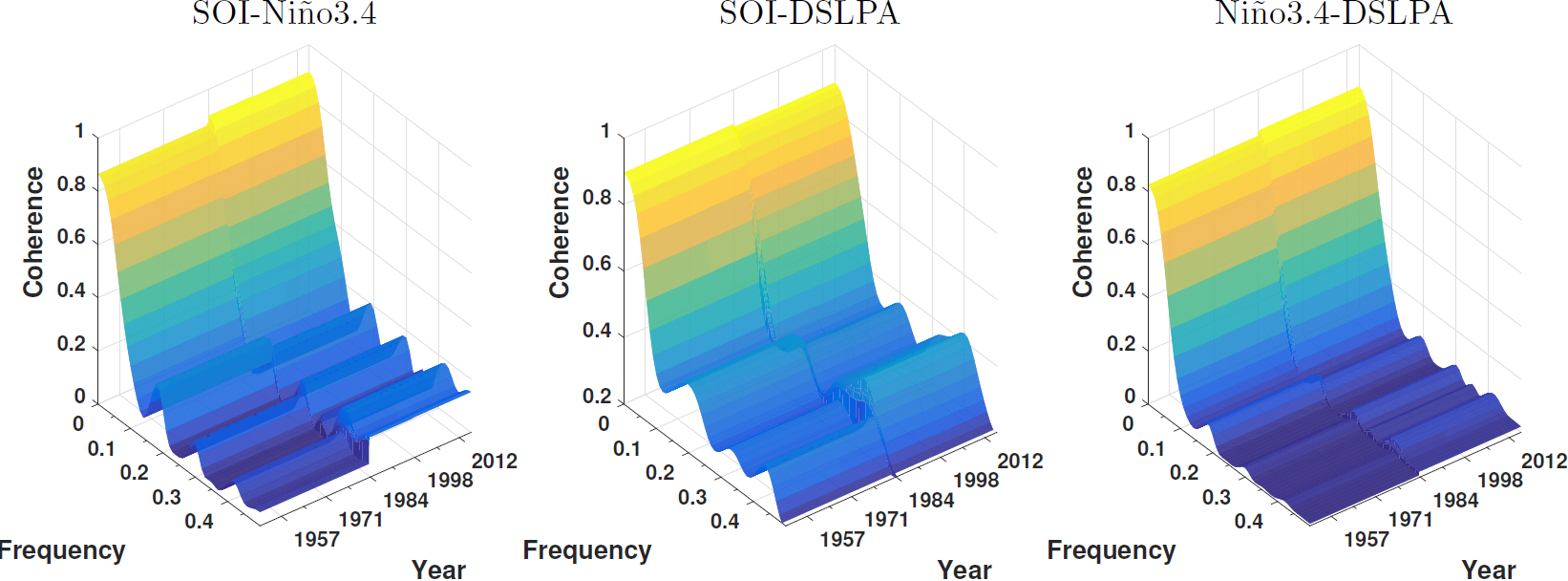}
		\end{minipage}
		\caption{Estimate of the time-varying log spectra and coherence of the SOI, Ni\~no3.4 and DSLPA series. }
		\label{spectsoi}
	\end{figure}

Plots of estimated time-varying log-spectra and pairwise coherences appear in Figure \ref{spectsoi}.  They indicate fairly pronounced changes in the spectrum of SOI and DSLPA, while the spectrum of Ni\~no 3.4 seems stationary across time. The spectra of SOI and DSLPA, and all coherences, show shifts in power from low frequencies to higher frequencies. Further, the integrated total power of SOI increased.  These are indicative of a higher rate of ENSO occurrences in recent years.

\begin{table}[t]
	\centering
	\caption{ \label{segment} Posterior probability of the number of segments of the ENSO series.}{%
		\begin{tabular}{l c c c c c c c} \hline \hline
			Number of segments ($m$)	&  1 & 2  & 3 & 4 \\ \hline
			Posterior probability  & 0.3286    & 0.6521      & 0.0096  & 0.0097\\ \hline
		\end{tabular}
	}\end{table}
	
    Previous analyses of ENSO have produced contradictory results.  In the geology literature, the analyses of \cite{solow2006} and \cite{nicholls2008} found ENSO to be stationary.  These results were further supported by the analysis of \cite{rosen2012}, who estimated that the posterior probabilities of stationarity to be 0.95, 0.93, and 0.99 for the SOI, Ni\~{n}o3.4, and DSLPA indices, respectively.  However, the analyses of \cite{trenberth1996} and \cite{timmermann2004} indicate a change point between 1980--1989, and the analyses of \cite{vecchi2010} and \cite{heureux2013} suggest that the  variability and amplitude in SOI has increased since 1980.  Our results are more consistent with these analyses that suggest changes in ENSO during the 1980's.  There are several possible reasons for the difference between our findings and those of \cite{rosen2012}.  First, we used data from 1951 to 2016, which contains five additional years of recent observation, and thus increases the possibility of detecting changes. Second, our method provides a joint analysis of the three series, taking into account possible changes in the coherence, while \cite{rosen2012} analyzed each series univariately. It is possible that the change in the SOI series is too subtle to be individually identified.  However, by simultaneously utilizing information in the coherence and individual spectra, our method could increase the power to detect smaller second-order changes.  Lastly, it should be noted that, although our analysis can be interpreted as suggesting a change in the ENSO spectrum during the 1980's, the posterior probability of stationarity was estimated as a nontrivial .332.  In this sense, our analysis is consistent with the geology literature as a whole, where several researchers have reported a change in ENSO during the 1980's, while others have reported ENSO being stationary over this period of time.
	
\section{Concluding Remarks} \label{remarks}
We conclude this article with a discussion of some limitations and possible extensions of the proposed methodology.
%
%
First, theoretically, the proposed methodology provides a means of conducting a time-varying spectrum analysis on an $N$-dimensional time series of general dimension $N$.  However, the number of Cholesky components grows quadratically with $N$, so that the proposed sampling scheme would be impractical for larger dimensions.   For instance, although the 2-channel EEG considered in Section \ref{sec:eeg} is standard in the analysis of sleep, in other domains, high-density 256-channel EEG are common.  A possible extension of the proposed method for the analysis of higher-dimension time series could be through the incorporation of time-dependent frequency domain principal components \citep{ombao2006} within the sampling algorithm.
Second, the proposed procedure is designed for the analysis of a single multivariate time series.  Many applications are concerned not with the analysis of a single multivariate time series, but of the analysis of replicated multivariate time series and in how their time-varying spectra are associated with other variables, such as clinical outcomes.   The method developed by \cite{fiecas2016} enables such an analysis when dynamics are smooth.  A possible extension of the proposed procedure to the replicated time series setting could involve simultaneously dividing the grid of time and clinical outcome values into approximately stationary blocks in a manner similar that which was recently proposed in the univariate setting by \cite{bruce2017}.
Third, the proposed approach provides an automated way to analyze time series with abrupt and smooth second--order changes. The smoothly-varying spectrum is recovered by averaging over the posterior distribution of partitions.  A challenging problem arises when a spectrum changes smoothly but rapidly over time.  The proposed procedure's ability to distinguish smooth but rapid changes from an single abrupt change point is limited by the sampling rate and care should be taken when interpreting the results in such scenarios.
Fourth, the Gaussian approximation (see Section 3.2) provides an efficient and flexible way to sample coefficients of basis functions when the number of observations in a stationary block is moderate or large \citep{le1953,le1990}. For a time series with small stationary blocks, more advanced distributional approximation approaches, such as variational Bayes and expectation propagation, could be adapted but with an increases in computational complexity.
Lastly, the procedure models the components of local spectra using penalized splines.  This approach assumes smoothness in spectra as functions of frequency, but other bases and regularizing priors could be used to adapt to specific features of a given problem.  For instance, Bayesian wavelet shrinkage \citep{chipman1997} could be used if near line spectra are expected or B-spline bases with knots selected to focus attention within certain frequency bands could be used if it is scientifically known a priori that power is contained within certain frequencies.




\section*{Supplementary Material}
Supplementary Material available online includes a pdf file that describes convergence assessment of the RJMCMC, further details of the analysis of the EEG data discussed in Section \ref{sec:eeg}, additional analyses of the ENSO data discussed in Section \ref{sec:enso}, and full details of the sampling scheme.   Both R and Matlab code for implementing the proposed procedure are provided.

\bibliographystyle{asa}
\bibliography{locstatbib}

\end{document}